\documentclass[twocolumn,showpacs,preprintnumbers,amsmath,amssymb]{revtex4}
\usepackage{graphicx}% Include figure files
\usepackage{dcolumn}% Align table columns on decimal point
\usepackage{bm}% bold math

\begin{document}

%\preprint{APS/123-QED}

\title{Observation of spin-wave propagation in permalloy microstripes}

\author{Andreas Krohn}
 \altaffiliation
{corresponding author email: andreas.krohn@physnet.uni-hamburg.de}
\affiliation{Institut f\"ur Angewandte Physik und
Mikrostrukturforschungszentrum, Universit\"at Hamburg,
Jungiusstrasse 11, D-20355 Hamburg, Germany}
\author{Sebastian Mansfeld}
\affiliation{Institut f\"ur Angewandte Physik und
Mikrostrukturforschungszentrum, Universit\"at Hamburg,
Jungiusstrasse 11, D-20355 Hamburg, Germany}
\author{Jan Podbielski}
\affiliation{Institut f\"ur Angewandte Physik und
Mikrostrukturforschungszentrum, Universit\"at Hamburg,
Jungiusstrasse 11, D-20355 Hamburg, Germany}
\author{Jesco Topp}
\affiliation{Institut f\"ur Angewandte Physik und
Mikrostrukturforschungszentrum, Universit\"at Hamburg,
Jungiusstrasse 11, D-20355 Hamburg, Germany}
\author{Wolfgang Hansen}
\affiliation{Institut f\"ur Angewandte Physik und
Mikrostrukturforschungszentrum, Universit\"at Hamburg,
Jungiusstrasse 11, D-20355 Hamburg, Germany}
\author{Detlef Heitmann}
\affiliation{Institut f\"ur Angewandte Physik und
Mikrostrukturforschungszentrum, Universit\"at Hamburg,
Jungiusstrasse 11, D-20355 Hamburg, Germany}
\author{Stefan Mendach}
\affiliation{Institut f\"ur Angewandte Physik und
Mikrostrukturforschungszentrum, Universit\"at Hamburg,
Jungiusstrasse 11, D-20355 Hamburg, Germany}

\date{\today}

%%%%%%ABSTRACT%%%%%%%%%%%%%%%%%%%%%%%%%%%%%%%%%%%%%%%%%%%%%%%%%%%%%
\begin{abstract}
We report on the propagation of spin waves in permalloy
microstripes. By means of scanning Kerr microscopy combined with
continuous microwave excitation, we detect the time evolution of
spin-wave interference patterns in an external magnetic field.
Assuming transverse spin-wave quantization we can directly measure
the amplitude, phase velocity and damping for the corresponding
transversal wave mode numbers $m$. We find that the spin-wave
interference pattern is dominated by $m=0$ and $m=2$ with phase
velocities $v_{\text{ph,0}}=71$~km/s and $v_{\text{ph,2}}=47$~km/s,
respectively.
\end{abstract}


\maketitle
%%%%%%INTRODCUTION%%%%%%%%%%%%%%%%%%%%%%%%%%%%%%%%%%%%%%%%%%%%%%%%%%%%%
Spin dynamics in patterned ferromagnetic elements have gained
growing interest since they exhibit fundamental magnetic phenomena.
The understanding of these principles is the key for technical
applications in the next decades e.g. data storage, sensors or
signal processing. Recently increasing interest has been devoted to
localization and propagation of spin waves in permalloy (Py) micro
and nano structures, e.g. wires \cite{Demidov2008,Demidov2008a},
arrays of wires and rings \cite{Topp2008,Giesen2005a}, films and
lattices \cite{Perzlmaier2008,Neusser2008}, or even complex 3D
shaped structures \cite{mendach:262501}.

%%%%%%Here we%%%%%%%%%%%%%%%%%%%%%%%%%%%%%%%%%%%%%%%%%%%%%%%%%%%%%
Demidov and coworkers \cite{Demidov2008a} recently observed a
complex magnetizatin pattern in transversally magnetized permalloy
microstripes employing time integrating micro-focus Brillouin light
scattering. They explained their observations by the interference of
spin-wave modes which are propagating along the stripe axis with
transverse quantization. In this letter we present time and
spatially resolved studies on transversally magnetized
Py~microstripes. By means of scanning Kerr microscopy combined with
microwave excitation (FMR-SKM)\cite{Tamaru2002,Neudecker2006} we
obtain experimental access to the time evolution of the spin-wave
interference pattern. Assuming spin-wave quantization transverse to
the Py~microstripes this data allows us to obtain separated images
of the interfering spin-wave modes. We can directly measure their
amplitude, phase velocity and damping. Our measurements agree well
with a dipole-based model for our stripe geometry
\cite{Guslienko2002}.

%%%%%%Preparation%%%%%%%%%%%%%%%%%%%%%%%%%%%%%%%%%%%%%%%%%%%%%%%%%%%%%
Our samples consist of a GaAs substrate with a 200-nm-thick coplanar
waveguide (CPW) fabricated on top. As shown in Fig.~1(a) the inner
conductor is 3~$\mu$m wide. The gaps between inner conductor and
ground lines are 2.5~$\mu$m wide. The CPW is planarized and
electrically insulated by a 250~nm thick layer of
hydrogensilsesquioxane (HSQ). With electron beam lithography and
thermal evaporation we place 20-nm-thick, 80~$\mu$m long and
5~$\mu$m wide $\text{Ni}_{80}\text{Fe}_{20}$ stripes on top of this
dielectric layer, crossing the CPW at right angle.

%%%%%%Setup%%%%%%%%%%%%%%%%%%%%%%%%%%%%%%%%%%%%%%%%%%%%%%%%%%%%%

We utilize the magneto-optic Kerr effect to detect the magnetization
dynamics. For this purpose we set up a time-resolved scanning Kerr
microscope with a spatial resolution better than 1~$\mu$m and a
temporal resolution better than 1 ps. We use a customized microwave
synthesizer (Parzich, ITS-9200) that can be locked to a laser-pulse
train with a repetition frequency between 74.1 and 77.9~MHz. This
circumvents the need of a stabilizing system for our pulsed laser
system (Coherent, Mira900f).
\begin{figure}[htbp]
\center
\includegraphics[width=8.5cm]{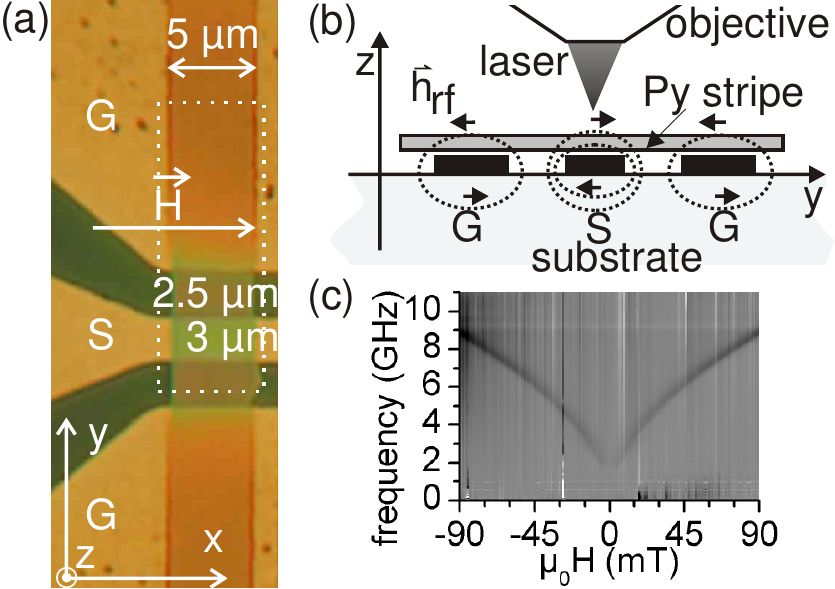}
\caption{\small{(color online) (a) Top view micrograph of the 20~nm
thick permalloy stripe prepared on top of the coplanar wave\-guide
(CPW). The Py~stripe is oriented perpendicular to the signal line
(S) and the two ground lines (G) of the CPW. The external field $H$
is applied in x direction perpendicular to the stripe axis. (b)
Spin-wave dispersion of the stripe obtained by microwave absorbtion
spectroscopy in Demon-Eschbach geometry (dark lines correspond to
strong absorption). (c) Schematic cross-section along the axis of
the Py~stripe perpendicular to the CPW plane. The high-frequency
current driven by the microwave produces a rf-magnetic field that
excites spin precession in the Py~stripes. The spin precession is
stroboscopically detected with focused fs laser pulses via the polar
magneto optical Kerr effect with a temporal resolution below 1~ps
and a spatial resolution below 1~$\mu$m.}} \label{1}
\end{figure}
The synthesizer generates microwaves
between 4 and 10~GHz with an increment corresponding to the laser
repetition rate of about 76~MHz. The phase between microwave and
laser pulses can be adjusted within $2 \pi$. In the following
experiments we saturate the sample at $\mu_0H=90$~mT perpendicular
to the stripe axis (Fig.~1(a)) and then decrease the external field
to the specified value. First, we determine the spin-wave eigenmode
spectrum of the Py~stripes by means of broadband microwave
spectroscopy (Fig.~1(c)). For a given resonance condition we then
pass the laser-locked microwave through the CPW to excite
magnetization dynamics in the Py~stripes and stroboscopically
measure the polar Kerr signal (Fig.~1(b)) \cite{Tamaru2002}. Spatial
resolution is obtained by scanning the sample under the focused
laser pulses.

%%%%%%Measurement%%%%%%%%%%%%%%%%%%%%%%%%%%%%%%%%%%%%%%%%%%%%%%%%%%%%%
\begin{figure}[htbp]
\center
\includegraphics[width=8.5cm]{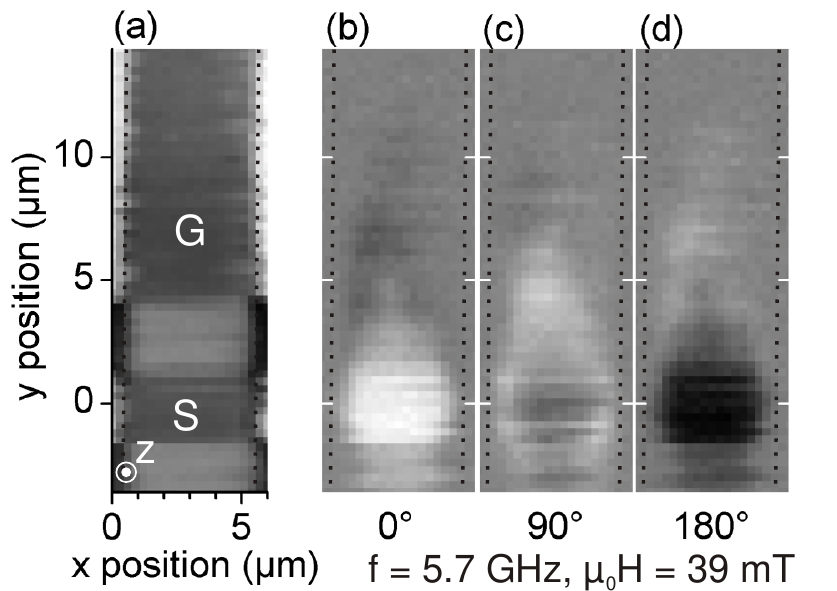}
\caption{\small{(a) Scanning reflection image of the sample area
marked by the dotted box in Fig.~1(a). Due to different reflectivity
of permalloy on gold (dark gray) and permalloy on gallium arsenide
(light gray)we can image the CPW below the Py~stripe. Pure GaAs and
Gold are coded black and white, respectively. (b)-(c)
Magnetic-contrast image of the stripe for three different phases
between the laser pulses and the microwave excitation. The edges of
the stripe are marked by a dotted line. Black and white correspond
to the maximum deflection of the spins in opposite directions. The
reflection and magnetic-contrast images are mapped simultaneously
enabling a precise spatial assignment of the spin-wave pattern.}}
\label{2}
\end{figure}In Fig.~2 we show data obtained at an excitation frequency of
5.7~GHz at an external field of 39~mT. Figure~2(a) shows a
reflection image of the sample that was taken simultaneously to the
magnetic-contrast images of the stripe depicted in Fig.~2(b)-(d). In
Fig.~2(a) one can clearly identify the signal and ground lines of
the CPW as well as the edges of the permalloy stripes (marked with
dotted lines). In the magnetic-contrast images we see the polar Kerr
rotation corresponding to the $z$ component of the magnetization
$m_z$ of the permalloy. Black and white indicate the maximum value
on opposite z directions. The three plots in Fig.~2(b)-(d) show
three phases of the microwave excitation relative to the probe-laser
pulses. In Fig.~2(b) we find a white area at the position of the
signal line of the CPW. When the phase $\phi$ between laser pulse
and microwave is changed by 90 degree (Fig.~2(c)), the white area
travels some micrometer along the stripe and changes its transversal
profile. At the position of the excitation we do not find a gray
area, as expected for a sinusoidal wave, but a pattern of black and
white areas. In Fig.~2(d) the phase is shifted another 90 degrees
and we observe a black area at the position of the excitation. The
white area has moved further on and lost intensity due to damping in
the material. In all three magnetic-contrast images we find a
vanishing amplitude of the magnetization precession at the edge of
the stripes which suggests a strong pinning of the magnetization at
these sites.

%%%%%%Measurement overview%%%%%%%%%%%%%%%%%%%%%%%%%%%%%%%%%%%%%%%%%%%%%%%%%
The transversal profile of the spin wave traveling along the stripe
reflects the confinement of the spin wave in the transversal
geometry. Following Ref.~\cite{Guslienko2003,Demidov2008a} we assume
quantization rules in the form $k_{x,m}=(m+1)\pi/w_{\text{eff}}$
where $m$ is the quantization number and $w_{\text{eff}}$ represents
the effective width of the stripe. Taking into account that we do
not excite modes with odd quantum numbers \cite{Kittel1958} and that
only the two strongest modes $m=0$ and $m=2$ are relevant we get the
general form of the magnetization:
\begin{align}
m_z&(x,y,t)\propto \nonumber\\
A_0\exp\left(\kappa_0 y\right)\cos&\left(k_{y,0}y-\omega\Delta
t+\varphi_0\right)\sin\left(\frac{\pi
x}{w_{\text{eff}}}\right) \nonumber\\
 +A_2\exp\left(\kappa_2 y\right)\cos&\left(k_{y,2}y-\omega\Delta
t+\varphi_2\right)\sin\left(\frac{3\pi x}{w_{\text{eff}}}\right).
\end{align}
\noindent Here $k_{y,0}$ and $k_{y,2}$ represent the wave vectors
along the stripe axis for the modes $m=0$ and $m=2$, $A_0$ and $A_2$
are the amplitudes of the modes. $\varphi_0$ and $\varphi_2$ take
into account the phases between the two modes and the microwave
excitation. $\kappa_0\;\text{and}\;\kappa_2$ are specific damping
parameters. $\omega=2\pi f$ is related to the excitation frequency
of the microwave. Since our magnetic-contrast images are taken along
the $x$ axis in subsequent line scans, we fit our data for fixed $y$
values. Figure~3(a) shows the magnetization $m_z$ in arbitrary units
for several time steps (dots) together with the fitted curves from
Eq.~(1) (lines) for $y=0$. As we know the frequency of the microwave
we can transform the phase shift $\Delta\phi$ between microwave and
laser pulses into a time interval $\Delta t=\Delta\phi/\omega$. With
the fitting parameters $A_0$, $A_2$, $(k_{y,0}\cdot y)$,
$(k_{y,2}\cdot y)$, $\varphi_0$ and $\varphi_2$ for all y positions
we can separate the two dominant modes.
\begin{figure}[htbp]
\center
\includegraphics[width=8.5cm]{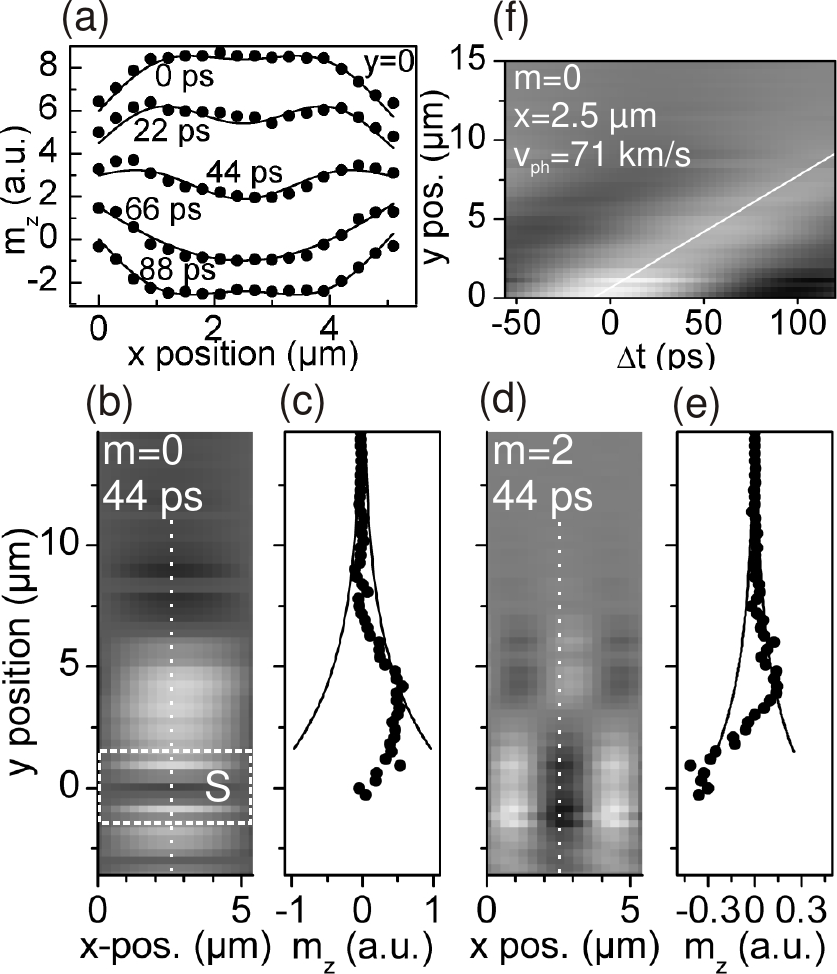}
\caption{\small{(a) $z$ component of the magnetization across the
Py~stripe at $y=0$ along the $x$ direction for five different phase
steps, i.e., points in time. Circles correspond to the Kerr signal,
lines correspond to fits with Eq. (1). (b),(d) Patterns of the
separated modes obtained from the fitting parameters ($A_0$, $A_2$,
$(k_{y,0}\cdot y)$, $(k_{y,2}\cdot y)$, $\varphi_0$ and $\varphi_2$)
at 44~ps. The signal line is marked with the white rectangle.
(c),(e) cross-section along the stripe (dotted lines in (b),(d)) and
damping curves. (f) Temporal evolution of the cross-section along
the stripe. The slope of the white area (white line) represents the
phase velocity of the mode and corresponds to a wave vector $k=0.5
\cdot 10^4~\text{cm}^{-1}$.}} \label{3}
\end{figure}
Comparing the amplitudes we find the higher mode to be four times
weaker than the fundamental mode which reflects the decrease in
oscillator strength of the confined spin wave with increasing mode
number \cite{Kittel1958}. The phase shift between the modes is
$|\varphi_0-\varphi_2|=128^{\circ}$~. Using these parameters, we can
construct 2D plots of the individual modes for every time step.
Figure~3(b) and (d) show the separated modes at $\Delta t=44$~ps. In
the area covering the signal line (white, dashed rectangle) the
magnetization is driven by the microwave. We extract the damping
parameters from the decrease of the transversally integrated
spin-wave intensity along the stripe, starting at the edge of the
signal line. With $\kappa_0=\kappa_2=-0.25~\mu\text{m}^{-1}$ we
receive identical values for the damping of both modes within the
accuracy of the measurement. Figure~3(c) and (d) show a
cross-section along the center of the stripe (dotted line in
Fig.~3(b)and(d))together with the corresponding damping curves. From
the time evolution of these plots we can directly measure the phase
velocity of the individual modes. To extract the phase velocity we
plot a cross-section along the center of the stripe (dotted line in
Fig.~3(b)) over time $\Delta t$ (Fig.~3(f)). From the slope of the
white area we obtain the phase velocity of the respective mode. For
the mode $m=0$ we determine a phase velocity of
$v_{\text{ph,0}}=71$~km/s and for the mode $m=2$ we find
$v_{\text{ph,2}}=47$~km/s corresponding to wave vectors
$k_0=0.5\cdot 10^{4}\text{cm}^{-1}$ and $k_2=0.77\cdot
10^{4}\text{cm}^{-1}$, respectively. This is in good agreement with
calculations on the dispersion relations of spin-wave modes based on
the dipole-exchange model \cite{Guslienko2003} for our stripe
geometry.

%%%%%%Conclusion%%%%%%%%%%%%%%%%%%%%%%%%%%%%%%%%%%%%%%%%%%%%%%%%%
In conclusion, we have shown time and spatially resolved
measurements of spin waves in transversally magnetized permalloy
stripes obtained with a combination of scanning Kerr microscopy and
continuous microwave excitation. Spin-wave confinement in the
stripes leads to spin-wave modes with different wave vectors that
travel along the stripe producing a characteristic interference
pattern. These experiments allow us to separate the individual
spin-wave modes in the time and spatial domain and to extract
amplitude, phase velocity, and damping of the respective modes. The
obtained data is in good agreement with a dipole-exchange based
model for the spin-wave dispersion in Py~microstripes.

This work was supported by the DFG via SFB 668, SFB 508 and GrK 1286.\\

\end{document}